\documentclass[
]{ceurart}

\usepackage[utf8]{inputenc}
\usepackage[english]{babel}
\usepackage{graphicx}
\usepackage{float}
\usepackage{hyperref}
\usepackage{array}
\usepackage{adjustbox}
\usepackage{geometry}
\usepackage{caption}
\usepackage{subcaption}

\usepackage{amssymb}

\begin{document}

\copyrightyear{2021}
\copyrightclause{Copyright for this paper by its authors.\\
  Use permitted under Creative Commons License Attribution 4.0
  International (CC BY 4.0).}

\conference{BIR~2021: 11th International Workshop on Bibliometric-enhanced Information Retrieval at ECIR~2021, April 1, 2021, online}

\title{Improving reference mining in patents with BERT}

\author[1]{Ken Voskuil}[%
email=k.s.voskuil@umail.leidenuniv.nl,
]
\author[1]{Suzan Verberne}[%
orcid=0000-0002-9609-9505,
email=s.verberne@liacs.leidenuniv.nl,
url=http://tmr.liacs.nl,
]
\address[1]{Leiden Institute of Advanced Computer Science, Leiden University}

\begin{abstract}
In this paper we address the challenge of extracting scientific references from patents. We approach the problem as a sequence labelling task and investigate the merits of BERT models to the extraction of these long sequences.
References in patents to scientific literature are relevant to study the connection between science and industry. Most prior work only uses the front-page citations for this analysis, which are provided in the metadata of patent archives. 
In this paper we build on prior work using Conditional Random Fields (CRF) and Flair for reference extraction. We improve the quality of the training data and train three BERT-based models on the labelled data (BERT, bioBERT, sciBERT).   
We find that the improved training data leads to a large improvement in the quality of the trained models. In addition, the BERT models beat CRF and Flair, with recall scores around 97\% obtained with cross validation. With the best model we label a large collection of 33 thousand patents, extract the citations, and match them to publications in the Web of Science database. We extract 50\% more references than with the old training data and methods: 735 thousand references in total.  
With these patent--publication links, follow-up research will further analyze which types of scientific work lead to inventions.

\end{abstract}


%
%
%
\maketitle              
\begin{abstract}
We

 \end{abstract}

\section{Introduction}
References in patents to scientific literature provide relevant information for studying the relation between science and technological inventions. These references allow us to answer questions about the types of scientific work that leads to inventions. Most prior work analysing the citations between patents and scientific publications focuses on the front-page citations, which are well structured and provided in the metadata of patent archives such as Google Patents. It has been argued that in-text references provide valuable information in addition to front-page references: they have little overlap with front-page references \cite{bryan2019text} and are a better indication of knowledge flow between science and patents \cite{nagaoka2015use,bryan2016impact,bryan2019text}.

In the 2019 paper by Verberne et al. \cite{Verberne2019ExtractingAM}, the authors evaluate two sequence labelling methods for extracting in-text references from patents: Conditional Random Fields (CRF) and Flair. In this paper we extend that work, by (1) improving the quality of the training data and (2) applying BERT models to the problem. We use error analysis throughout our work to find problems in the dataset, improve our models and analyze the types of errors different models are susceptible to.

We first discuss the prior work in Section~\ref{sec:priorwork}. We describe the improvements we make in the dataset in Section~\ref{sec:data}, and the new models proposed for this task in Section~\ref{sec:bert}. We compare the results of our new models with previous results, both on the labelled dataset and a larger unlabelled corpus (Section~\ref{sec:results}). We end with a discussion on the characteristics of the results of our new models (Section~\ref{sec:discussion}), followed by a conclusion.

Our code and improved dataset are released under an open-source license on github.\footnote{\url{https://github.com/kaesve/patent-citation-extraction}}


\section{Prior work} \label{sec:priorwork}

Reference analysis in patents has primarily been done using the references that are listed on the patent's front page. Patents often contain many more references in the patent text themselves, but these are more difficult to extract and analyze because their formatting is not standardized. Verberne et al.~\cite{Verberne2019ExtractingAM} introduce a new labelled dataset consisting of 22 patents and 1,952 hand-labelled references. They apply two sequence labelling methods to the reference extraction tasks.

Conditional Random Fields (CRF) model sequence labelling problems as an undirected graph of observed and hidden variables, to find an optimal sequence of hidden variables (labels) given a sequence of feature vectors \cite{wallach2004conditional}. Feature vectors usually consist of several manually designed heuristics on the level of individual tokens and small neighborhoods of tokens. For extracting references, Verberne et al. \cite{Verberne2019ExtractingAM} use a set of $11 + 6*4$ features. This includes 11 features derived from the current token, ranging from the part-of-speech (POS) tag (extracted with NTLK), lexical features such as whether the token starts with a capital or is a number, and pattern-based features to mark tokens that look like a year or a page number.\footnote{The features are similar to the ones used in \url{https://sklearn-crfsuite.readthedocs.io/en/latest/tutorial.html}} It also includes a subset of 6 features for each of the two preceding and following tokens.

As the authors note, CRF has limited capabilities to take context into account. They chose to compare CRF with the Flair framework, which is better able to use token contexts. Flair uses a BiLSTM-CRF model in combination with pre-trained word embeddings \cite{akbik2019flair}. One downside of Flair models is that they are memory intensive, which limits the maximum sequence length it can process at once. Where the CRF model can analyze a complete patent at once, the Flair models required to split sequences up into subsequences of 20 to 40 tokens \cite{Verberne2019ExtractingAM}. Verberne et al. used the IOB labels during training to prevent splitting within a reference.

The models were evaluated by measuring precision and recall using cross validation on the labelled data. CRF performed better than Flair in all measures except the recall of I-labels. The models were also applied to a large corpus of 33,338 unlabelled USPTO biotech patents, and the resulting extracted references were matched against the Web of Science (WoS) database. Here, Flair performed significantly better. Counting references with a definitive match in WoS that were not included in the patent front-page, CRF was able to find 125,631 of such references compared to 493,583 references found by Flair.

Recent developments in transfer learning have improved the state of the art in numerous NLP tasks. BERT~\cite{devlin2019bert} is a large transformer model that is \textit{pre-trained} on a large corpus for multiple language modelling tasks. The resulting model can be used as a basis for new tasks on different data sets. Even when the contents of these data sets or the task deviate significantly from the pre-training corpus and tasks, the pre-training is still beneficial. Several authors have trained models with the same architecture as BERT on different, more domain-specific corpora. These include SciBERT~\cite{DBLP:journals/corr/abs-1903-10676} and BioBERT~\cite{DBLP:journals/corr/abs-1901-08746}.

\section{Improving data quality} \label{sec:data}

While exploring the results of our models, we found that several prediction errors seemed to be caused by mistakes in the labelled data. These mistakes result in a more pessimistic evaluation of our models and, more importantly, could influence the effectiveness of training our models. We noticed two types of problems; inconsistent or missing labels, and inconsistent tokenization. We include examples of both kinds of problems below, and describe our attempts to improve the data quality.

\subsection{pre-processor inconsistencies}

The patent dataset contains text from 22 patents taken from Google Patents. Labels were added manually by one annotator using the BRAT annotation tool\footnote{\url{http://brat.nlplab.org/}}, and the text was subsequently transformed into IOB files using a pipeline consisting of splitting the text into sentences, then tokens and adding IOB and POS tags. Because tokenization was applied after annotation, the labels produced by BRAT needed to be aligned with the produced tokens. In some cases, this was done by recombining tokens. When comparing the source text with the IOB data, we found that some sequences of tokens seemed to have been accidentally reordered. An example of this is shown in Figure \ref{fig:examples} After reviewing the pre-processing pipeline we were able to find the likely cause of this problem. We chose to replace this pipeline with a simpler procedure, that does not do sentence splitting or combining of tokens. Besides sentence boundaries, our method also ignores paragraph boundaries and white space in general.

\begin{figure}
    \centering
    \subcaptionbox{Original text\break}{
        \begin{tabular}{l}
             (Eskildsen et al., Nuc. Acids Res. 31:3166-3173, 2003; \\
             Kakuta et al., J. Interferon \& Cytokine Res. 22:981-993, 2002.)
        \end{tabular}
    }
    \subcaptionbox{Original tokenization\break}{
        \begin{tabular}{c|cccccccccc}
             Token & Eskildsen & et & al., & Nuc. & Acids & Res. &  \dots & Res. & 22:981-993, & \textbf{2002.)(} \\
             Label & B & I & I & I & I & I & \dots & I & I & \textbf{O}
        \end{tabular}
    }
    \subcaptionbox{New tokenization}{
        \begin{tabular}{c|cccccccccccccccc}
             Token & ( & Eskildsen & et & al. & , & Nuc & . & Acids & \dots & Res & . & 22:981-993 & , & 2002 & . & ) \\
             Label & O & B & I & I & I & I & I & I & \dots & I & I & I & I & I & I & O
        \end{tabular}
    }
    \caption{Comparing the original with the new tokenization. Note that punctuation marks are now treated as separate tokens. Also note the labelling of the braces, and the wrongly labelled last token in the original tokenization. Example taken from patent US8133710B2.}
    \label{fig:examples}
\end{figure}

\subsection{Inconsistent labelling}

After improving the pre-processing, we still found examples of label inconsistencies. Moreover, our models found several references that were not included in the annotations. Finally, we found multiple instances of references to patents and other non-academic literature. These are often hard to distinguish from scientific literature references. We manually looked at each difference between predicted and expected labels, and changed the annotations where necessary. We repeated this process several times, with different models and after retraining on the updated data. In this process, we labelled 330 new references, resulting in a total of 2,318 references and 32,359 (I)nside tokens. We chose to include patent references when they included author names or titles, and other non-literature references, when the reference shares the format of a literature reference. This simplifies the task, as the model does not have to disambiguate references by their type. Since these extracted non-literature references will not match with the publications in WoS, they will be filtered out in the next step of the pipeline.

While we think this process has improved the data quality significantly, our method does introduce biases in the training and evaluation of our models. By only fixing labelling mistakes that our models find, we may overlook unlabelled references that our models miss. This leads to an overestimation in our evaluation, and biases in our model due to the feed back loop in the training process. By using multiple different models for finding incorrect labels, we mitigate the effect to some extent. Beside an intrinsic evaluation using the labelled data, we will also evaluate our models on an extrinsic task using unlabelled data. This allows us to still compare our model performance with previous results, without biases in the dataset or overestimations.

\section{Extracting references with BERT, BioBERT and SciBERT }
\label{sec:bert}

We compare three different pre-trained models for extracting references from our data set; BERT, BioBERT and SciBERT. 
Since our data set consists of patents from the biomedical domain, we expect that these more domain-specific pre-training corpora will have a positive effect on our task. Before comparing the results between these models, we describe our method for fine tuning the pre-trained model for reference extraction.


\subsection{pre-processing}

BERT-based models have two characteristics that require additional pre-processing of our dataset. BERT uses its own neural subword tokenization. Our dataset is already tokenized into words, as described above, so we apply the BERT tokenizer to each token in our dataset. Transformer-based models such as BERT also work on fixed sequence lengths, using padding for shorter sequences, and are memory intensive. The models we train use a maximum sequence length of 64 tokens, limited by the memory available. Though this can be configured to be higher depending on the available hardware and the size of the model, it is infeasible to apply these models on complete patents, which can contain tens of thousands of subword tokens. There are several common strategies to divide text into shorter sequences. A natural approach is to use paragraph or sentence splitting. We found this insufficient, as many sentences in our data set run for much longer than the limit of 64 tokens. Our data set contains not only long sentences; even references, the entities we are looking to extract, can be longer than 64 tokens. Because of this observation we decide to not use any semantic or structural information in splitting our text, except for our original token boundaries.

Our BERT specific pre-processing can be summarized in the following steps:

\begin{enumerate}
    \item Collect the sequence of tokens $T$ and their respective labels $L$ for a given patent
    \item Create two empty lists $T'$ and $L'$
    \item Add the sequence start token to $T'$
    \item While there are tokens left in $T$:
    \begin{enumerate}
        \item Get the next token $t$ and label $l$
        \item Use the word tokenizer to get sub tokens $t'_1, ..., t'_n$
        \item If $|T'| + n + 1$ is larger than our limit of 64 tokens or when we reach the end of the document, add the sequence end token to $T'$, pad both sequences and add them to the data set. Set $T'$ and $L'$ to new empty lists
        \item Add $t'_1, ..., t'_n$ to $T'$, add $l$ to $L'$
    \end{enumerate}
\end{enumerate}

We note that the retokenization changes our task from a one-to-one to a many-to-many sequence-to-sequence task, as there could now be multiple subword tokens associated with one label. Another implication of these pre-processing steps, is that the entities that we seek to extract can be split across multiple sequences of 64 subword tokens. As mentioned earlier, we have a total of 2,318 references and 32,359 tokens labelled as (I)nside. This gives us a total of $34,677$ reference tokens (labelled either B or I). We find that the average reference contains $\frac{34,677}{2,318}\approx15$ word tokens, and thus at least that many subword tokens. We can expect a large number of references to be split across two or more sequences. We expect that this could have a significant effect on the performance of our models, as the model will not always have access to the context of a reference.

\subsection{Training the BERT models}
We fine-tunet three different BERT models to our labelled data: BERT-base, bioBERT, and sciBERT (all cased). We used To fine-tune the BERT models, we use the open source BERT implementation by HuggingFace\footnote{\url{https://huggingface.co/transformers/model_doc/bert.html##bertfortokenclassification}}, with a token classification head consisting of a single linear layer. In the case that an input sequence is shorter than 64 tokens (which only occurs at the end of a patent), we mask out the loss for the output past the input sequence. We train the models for three epochs through our training data, with a batch size of 32.\footnote{We published the trained models on \url{https://github.com/kaesve/patent-citation-extraction}}

\section{Results} \label{sec:results}
\subsection{Intrinsic evaluation}\label{sec:intrinsic}
We evaluate our models using a leave-one-out training scheme. For each patent in the data set we train a new model using the other 21 patents as the training data. Aside  from  the  maximum  sequence  length,  we  used  the  default  hyperparameter configurations provided by the chosen framework. We evaluate on both the original and updated dataset.

\begin{table}[t]
    \caption{Comparing three BERT models and two baseline models using leave-one-out evaluation on 22 patents (micro averages). Best results are printed in boldface. * indicates results as reported by \cite{Verberne2019ExtractingAM}.  }

    \centering
    \begin{tabular}{r|c|c|m{1.4cm}m{1.4cm}m{1.4cm}r}
    \hline
                & Dataset  & Label & Precision & Recall &   F1 & Support \\ \hline \hline
           BERT & Original & B &      0.849 &   0.927 & 0.886 &  1,988 \\
                &          & I &      0.896 &   0.955 & 0.924 & 28,449 \\ \hline
        SciBERT & Original & B &      0.865 &   0.925 & 0.894 &  1,988 \\
                &          & I &      0.898 &   0.944 & 0.920 & 28,449 \\ \hline
        BioBERT & Original & B &      0.843 &   0.929 & 0.884 &  1,988 \\
                &          & I &      0.894 &   0.952 & 0.922 & 28,449 \\ \hline \hline
           BERT & Updated  & B &      0.934 &   0.948 & 0.941 &  2,318 \\
                &          & I &      0.985 &   0.972 & 0.978 & 32,359 \\ \hline
        SciBERT & Updated  & B & \textbf{0.947} &   0.954 & 0.950 &  2,318 \\
                &          & I & \textbf{0.986} & \textbf{0.976} & \textbf{0.981} & 32,359 \\ \hline
        BioBERT & Updated  & B &      0.944 & \textbf{0.957} & \textbf{0.951} &  2,318 \\
                &          & I & \textbf{0.986} &   0.974 & 0.980 & 32,359 \\ \hline \hline
            CRF* & Original & B & 0.890 & 0.824 & 0.856 &  1,988 \\
                &          & I & 0.914 & 0.870 & 0.891 & 28,449 \\ \hline
            CRF & Updated  & B & 0.922 & 0.893 & 0.907 &  2,318 \\
                &          & I & 0.964 & 0.938 & 0.951 & 32,359 \\ \hline
          Flair* & Original & B & 0.762 & 0.702 & 0.731 &  1,988 \\
        (Flair embeddings) & & I & 0.814 & 0.890 & 0.850 & 28,449 \\ \hline
          Flair* & Original & B & 0.722 & 0.647 & 0.682 &  1,988 \\
        (Glove embeddings) & & I & 0.789 & 0.840 & 0.814 & 28,449 \\ \hline
    \end{tabular}
    \label{tab:intrinsic_results}
\end{table}
Table \ref{tab:intrinsic_results} shows the results of evaluating the models on the labelled data using leave-one-out validation. We also include the results of \cite{Verberne2019ExtractingAM} as a baseline, however, the results are not directly comparable as they used five-fold cross-validation for evaluation. Their models therefore were trained on less data. Finally, we include the results of applying the original CRF implementation on our updated dataset, using the same leave-one-out validation strategy.

We see that our new models perform reasonably well on the original dataset. Comparing to the baseline methods, we see that the BERT models consistently achieve a much higher recall. This is especially useful for the WoS matching task, as was discussed earlier. 

When we compare the results of our models obtained with the updated dataset to those obtained with the original data, we see that the changes in the dataset lead to improvements in every metric. Especially in the precision column, we see a large jump in quality. This jump is in part the direct result of our relabelling process. Most changes in the dataset concerned changing labels from `O' to `I' or `B' tokens, where our models found references that were missed during labelling.

Comparing the BERT-based models with each other, we find that the differences are small. With the updated data the SciBERT and BioBERT models seem to perform slightly better than the plain BERT model.

Finally, we can compare the results of the CRF model on the original and updated dataset. We again see a clear jump in performance. This comparison does suffer from the training bias and different evaluation strategy mentioned earlier. Furthermore, the CRF model uses features designed for the original dataset. As we changed the tokenization process, this means that some of the pattern based features do not work as intended. Still, we think the results do show that the changes to the dataset make this task easier.

\subsection{Extrinsic evaluation}\label{sec:extrinsic}

We also apply each model to an unlabelled data set of 33,338 patents~\cite{Verberne2019ExtractingAM}. For this application, the models are trained on the complete labelled data set. The references produced by these models are matched against the Web of Science database, using the same procedures as reported in \cite{Verberne2019ExtractingAM}.

From the set of 33,338 patents, we extract references to papers published in the years 1980--2010 (the `focus years'). This results in a list of extracted references. We parse them into separate fields: first author, second author, year, journal title, volume/issue, and page numbers. Then we match those fields to publications in the database. If we find a non-ambiguous match for a subset of the fields, we count this as a `definite match'~\cite{Verberne2019ExtractingAM}.

\begin{figure}[t]
    \centering
    \includegraphics[width=1\columnwidth]{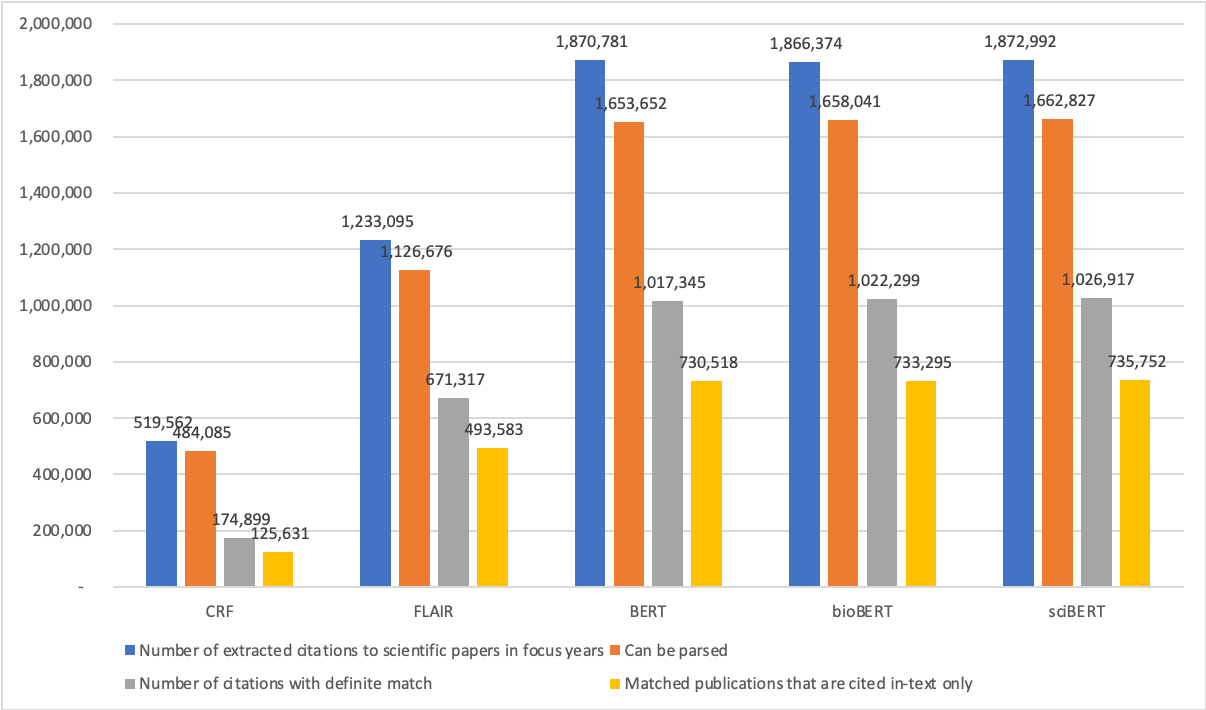}
    \caption{The number of found, parsed and matched references in the unlabelled dataset. The results for CRF and Flair were obtained for the original dataset. In-text citations are citations that only occur in the text and are not included in the patent front page. The `focus years' are the years 1980--2010, for which we extract references. A `definite match' is a non-ambiguously matched publication to the reference. (also see~\cite{Verberne2019ExtractingAM} for details about the matching procedure)}
    \label{fig:extrinsic_results}
\end{figure}

The results are displayed in Figure \ref{fig:extrinsic_results}. There is a clear difference between the new BERT-based models and the previous CRF and Flair models, but these results are not directly comparable since CRF and Flair were trained on the original data. The figure also shows that the three BERT models perform nearly identical to each other. As with the results from our intrinsic evaluation, SciBERT seems to perform better than the other two BERT models by a small margin.

We found that our models do not always produce clean sequences of IOB tokens; sometimes the beginning is not marked as a B, or a word in the middle of a reference is labelled as O. We extract references from sequences of I tokens starting with a B token or an I token preceded by an O token, and ending before an O or B token. In the case that our model misses a word in the middle of a reference, this means that we split this reference in two references during extraction. Our matching script reports unique matches per patent, so this does not lead to double-counting references. On the other hand, it could mean that neither part of the split reference contains enough information to make a definite match in the WoS database.

\section{Discussion} \label{sec:discussion}

Our results show that our BERT-based models outperform both CRF and Flair, especially after improving the training data. While the increased precision and recall is likely overestimated in our intrinsic evaluation, the new models also perform better in our extrinsic evaluation, which does not have the same training biases. Our models were able to extract roughly 240,000 more references that could be matched with the WoS database from the unlabelled data than Flair could, an increase of almost 50\%. 

The difference between the numbers of matched publications found by CRF and BERT is striking given the small differences in quality of the models measured with leave-one-out validation (Table~\ref{tab:intrinsic_results}). This can for a large part be explained by the improved training data, but also by the higher recall for the BERT models. In addition, we investigate two characteristics of errors made by our models, and show the differences between BERT and CRF. We focus on prediction errors \emph{within} references, as these have the largest effect on the downstream task of parsing references. Specifically, we look at cases where the model labels a token as O when that token is labelled as B or I in the ground truth.

\begin{figure}[t]
    \centering
    \includegraphics[width=\textwidth,trim={1cm 1cm 1cm 1cm },clip]{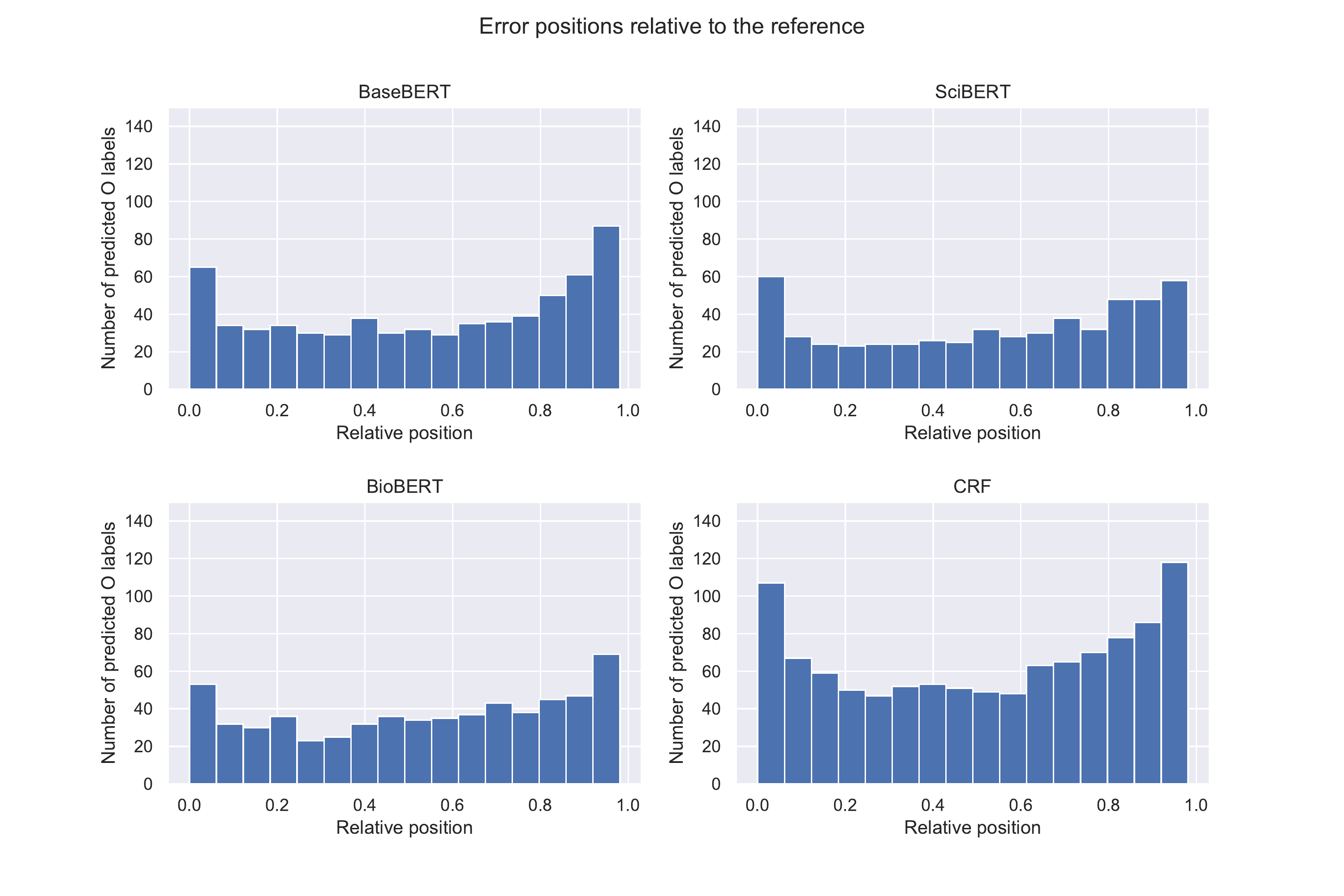}
    \caption{The relative positions of predicted `O' labels within references, per model.}
    \label{fig:relative_error_positions}
\end{figure}

Figure \ref{fig:relative_error_positions} shows the relative position of errors within references. This data was captured during the leave-one-out evaluation. One major difference between BERT and CRF-based models is that CRF explicitly learns ordered patterns in sequences. We would expect CRF models to make errors by starting or ending a label sequence too early or too late, but we do not usually expect errors to occur in the middle of a reference, as CRF learns that an I never follows an O. Without this structural prior, we expect the errors to occur more uniformly across the references for the BERT models. The histograms seem to confirm these intuitions. Leaving out mistakes in the first word, we see that the distribution for especially the SciBERT and BioBERT models seem uniform. The CRF model shows a clear drop in the first third of the distribution, and a steady increase in the second half. 

By manually looking at references where CRF predicts an O close to the middle, we found we could categorize these mistakes almost completely in two groups: CRF only labelled the first or last few tokens as part of the reference, or the reference is very long and CRF finds two references at beginning and end of the reference. In both scenarios CRF does produce coherent sequences of a B label followed by I labels. On the other hand, our BERT models sometimes do not predict a B at all, or in the wrong place. The models are also prone to missing an I label in the middle of a reference.

\begin{figure}
    \centering
    \includegraphics[width=0.8\textwidth]{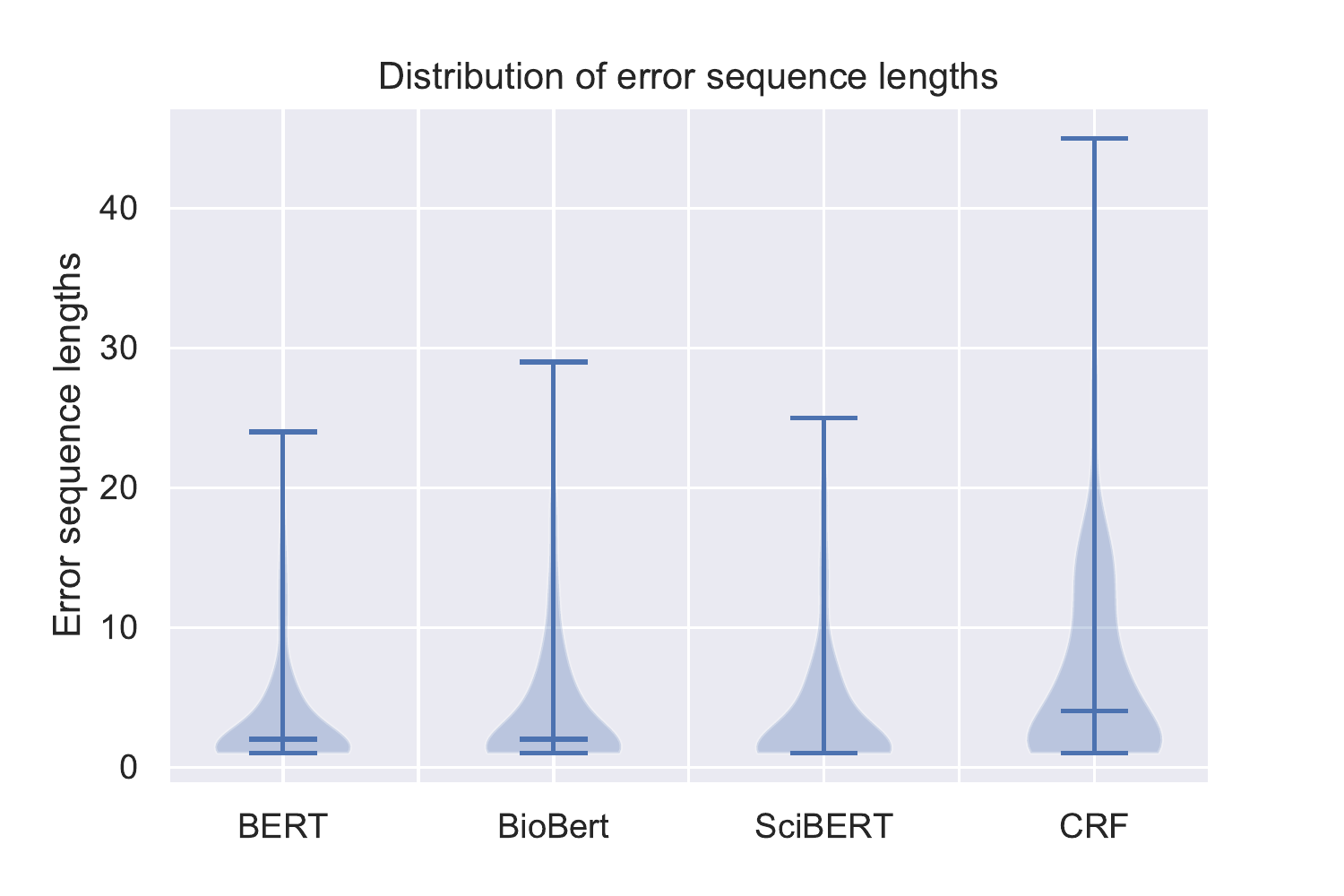}
    \caption{The relative positions of predicted `O' labels within references, per model.}
    \label{fig:error_length_violins}
\end{figure}

Figure \ref{fig:error_length_violins} is another way to visualize this difference. Here we plot the lengths of sequences of O's found within references. The median error sequence length is one or two for the BERT models, and four for CRF. In other words, BERT models not only make fewer mistakes than CRF, but the mistakes are smaller on average, and more uniformly spread across the reference. We speculate that this helps with the ultimate task of parsing and matching the references. CRF errors almost always include the first or last few tokens, which often contain important information for parsing the reference, such as the publication year and the author names.

\section{Conclusion}

We applied BERT-based models to extract references from patent texts. We found that these models achieve better recall than CRF and Flair. We use an external database of publications to match these references, which means that recall is more important than precision, as imprecisions will be resolved during matching. During the development of our models, we found that the original dataset for this task had errors in labelling and pre-processing. We used our models interactively to find these mistakes, and repaired them. 

We find that the improved training data leads to a large improvement in the quality of the trained models. In addition, the BERT models beat CRF and Flair, with recall scores around 97\% obtained with cross validation. Our models were also applied to a large unlabelled dataset, and were able to extract 50\% more references than previous methods.

We also show that BERT models are prone to a different kind of errors than CRF models. Combining these methods could potentially lead to a stronger model. We think that the limited maximal sequence size that BERT can handle affects its performance, due to the average length of references. Recent work focuses on modifying the attention architecture underlying BERT to better accommodate longer sequences. This includes new models such as the Reformer, Longformer, Linformer, Big Bird and the Performer \cite{tay2020efficient}. We think these models could achieve even better results, with little modification to our method.




\bibliography{bert_citations.bib}

\end{document}